\documentclass[prd,aps,twocolumn,superscriptaddress,showpacs,amssymb,floatfix]{revtex4}
\usepackage{epsfig}
\begin{document}

\def\Tr                 {\mathop{\rm Tr}}
\def\none               {\multicolumn{2}{c|}{---}}
\newcommand{\half}{\frac{1}{2}}
\newcommand{\spac}{\mbox{\hspace{2em}}}
\newcommand{\vev}{\frac{v}{g_3}}

\def\ph{\hat\Phi}
\def\ch{\hat\chi}
\def\U{{\mathbb I}}
\def\tch{\tilde\chi}
\def\tU{\tilde U}
\def\tV{\tilde V}
\def\tL{\tilde\Lambda}
\def\x{\vec{x}}
\def\xpmu{\x\!+\!\hat\mu}
\def\xpi{\x\!+\!\hat\imath}
\def\xpj{\x\!+\!\hat\jmath}
\def\xpij{\x\!+\!\hat\imath\!+\!\hat\jmath}
\def\tr{{\rm Tr}\,}
\def\nn{\nonumber}
\def\kh{\hat{k}}
\def\jh{\hat\jmath}
\def\ih{\hat\imath}
\def\xh{\hat{x}}
\def\yh{\hat{y}}
\def\zh{\hat{z}}
\def\O{{\mathcal O}}
\def\ie{{\it i.e.}}

\title{The monopole mass in the three-dimensional Georgi--Glashow
model}

\author{A.C. \surname{Davis}}
\author{A. \surname{Hart}}
\affiliation{DAMTP, CMS, University of Cambridge, Wilberforce Road,
Cambridge CB3 0WA, U.K.}

\author{T.W.B. \surname{Kibble}}
\affiliation{Blackett Laboratory, Imperial College, London SW7 2BW,
U.K.}

\author{A. \surname{Rajantie}}
\affiliation{DAMTP, CMS, University of Cambridge, Wilberforce Road,
Cambridge CB3 0WA, U.K.}

\begin{abstract}
We study the three-dimensional Georgi--Glashow model
to demonstrate how magnetic monopoles can be studied fully
non-perturbatively in lattice Monte Carlo simulations, without
any assumptions about the smoothness of the field configurations.
We examine the apparent contradiction between the
conjectured analytic connection
of the `broken' and `symmetric' phases, and the interpretation of
the mass (\ie, the free energy) 
of the fully quantised 't~Hooft--Polyakov monopole as an
order parameter to distinguish the phases. We use Monte Carlo
simulations to measure the monopole free
energy and its first derivative with respect to the scalar mass. On
small volumes we compare this to semi--classical predictions for the
monopole. On large volumes we show that 
the free energy is screened to
zero, signalling the formation of a confining monopole
condensate. This screening does not allow the monopole mass to be
interpreted as an order parameter, resolving the paradox.
\end{abstract}



\maketitle

\section{Introduction}

On the level of classical field equations, the three--dimensional
Georgi--Glashow model has two phases: When the mass parameter of the
Higgs field is negative, the SU(2) gauge symmetry is broken into U(1),
and when it is positive the symmetry is unbroken. The phase of the
system can be determined by a local measurement of, say, the scalar
field ${\rm Tr}\Phi^2$, which vanishes in the symmetric phase but is
non-zero in the broken phase.

In the broken phase, the field equations have a topologically
non-trivial solution, the 't~Hooft--Polyakov
monopole~\cite{'tHooft:1974qc,Polyakov:1974ek}, whose energy is
concentrated around a point-like core. The mass, \ie, the total energy
carried by a monopole, decreases when the mass parameter approaches
zero from below, and vanishes in the symmetric phase, in the sense
that the solution is indistinguishable from the trivial vacuum
solution.

In many cases, however, we are more interested in the behaviour of the
model when fluctuations are taken into account. It is immaterial
whether the fluctuations are thermal fluctuations in a classical field
theory or quantum fluctuations in a Wick--rotated (2+1)--dimensional
quantum field theory. Both of these systems are described by the same
partition function, and we shall make no distinction between
them. Nevertheless, we shall call the treatment based on classical
field equations ``semiclassical'' even though it is no more accurate
in a classical field theory at a non-zero temperature than it is in a
quantum field theory.

When fluctuations are present, the above simple picture changes
completely. In particular, the `symmetric' and `broken' phases are
believed to be analytically connected to each
other~\cite{Fradkin:1979dv,Nadkarni:1988pb,Hart:1997ac,Kajantie:1997tt}.
Order parameter candidates that are not gauge invariant, such as
$\langle \Phi \rangle$, vanish in both phases, and positive definite
observables, such as ${\rm Tr}\Phi^2$ mentioned above, are non-zero in
both phases. It would seem natural that a quantity like the mass of a
't~Hooft--Polyakov monopole, however, should be protected against the
effects of the fluctuations by its topology, and that it should
therefore serve as an order parameter for the phase transition. If
this were the case, the phases could not be analytically
connected. One example of this is the Abelian Higgs model, in which
the vortex tension indeed acts as an order
parameter~\cite{Kleinert:1982dz,Kovner:1991pz,Kajantie:1999zn}.

On the other hand, it is not even obvious that the monopole mass can
be given a rigorous definition in a fluctuating theory, because, in
general, one cannot assume that the field configurations that
contribute to the partition function are in some sense close to
solutions of classical field equations. This problem was solved in
Ref.~\cite{Davis:2000kv}, however, where the monopole mass was defined
as the increase of the free energy when the total magnetic charge of
the system is increased by one. Furthermore, it was shown how this
quantity can be measured in Monte Carlo simulations.

Thus, we have a well defined observable, the monopole mass, which
could naturally be expected to be zero in the symmetric phase and
non-zero in the broken phase, and still the phases are believed to be
analytically connected. The purpose of this paper is to explain this
apparent paradox.

First, we present a calculation based on a simple dilute monopole gas
approximation, which predicts that although the monopole free energy
is indeed non-zero and roughly equal to its classical value in a system
of intermediate volume, it decays to zero at exponentially large
volumes. Therefore, it should actually vanish everywhere in the
thermodynamic limit. This calculation is very similar to
Polyakov's argument~\cite{Polyakov:1977fu,polyakov87} 
that the photon has an exponentially small mass in
the broken phase.

Second, we measure the monopole free energy directly in a Monte Carlo
simulation on different volumes using the method developed in
Ref.~\cite{Davis:2000kv}. We find that the monopole free energy has a
volume-independent value in a wide range of lattice sizes, which shows
that it corresponds to a localised, point-like object. In agreement
with the analytical arguments, however, it eventually starts to
decrease, when the volume is large enough.

The vanishing of the monopole free energy in the infinite volume limit
implies that the monopoles condense.  This leads to confinement of
electric charge according to the dual superconductor
picture~\cite{Mandelstam:1976pi}, and our results can therefore be
considered as a numerical verification of Polyakov's semi-classical
argument \cite{polyakov87} that the Higgs phase is confining. In
particular, since the monopole free energy vanishes in both phases in
the infinite volume limit, it does not act as an order parameter, and
this resolves the apparent paradox between a smooth crossover and the
non-analytic behaviour of the monopole mass in the semiclassical
approximation.

Within the framework of high--temperature dimensional reduction
\cite{Kajantie:1996dw}, the three--dimensional Georgi--Glashow model
is an effective theory for the Yang--Mills theory at high temperatures
(see, for example,
\cite{hart99}
and references therein). The phase transition of our model, however,
is not related to the deconfinement phase transition of the
Yang--Mills theory or QCD. On the other hand, our methods can be
generalised to four dimensions in a straightforward way, and they may
therefore be applicable also to studying Abelian monopoles
\cite{'tHooft:1981ht} in the Yang--Mills theory, in particular whether
they condense at the transition point as has been suggested as a
possible ``mechanism'' for confinement \cite{Mandelstam:1976pi}.

Monopole free energies in the Yang--Mills theory have been studied
before by several groups~%
\cite{Smit:1994vt,DelDebbio:1995sx,Frohlich:1999wq,Cea:2000zr} using
different techniques.  In Refs.~\cite{Smit:1994vt,Cea:2000zr} fixed
boundary conditions were used to create a monopole, but this leads to
significant boundary effects.  In
Refs.~\cite{DelDebbio:1995sx,Frohlich:1999wq} a monopole creation
operator was used, which lets one measure not only the mass but also
correlation functions of the monopole field.  With periodic boundary
conditions, however, the operator creates not only a monopole, but
also an antimonopole somewhere in the system in order to satisfy
Gauss's law.  The advantage of our approach is that the system really
has a non-zero magnetic charge, and because translation invariance is
preserved, no singularities can arise even near the boundaries of the
lattice.

The structure of the paper is as follows.  We start by discussing the
three--dimensional Georgi--Glashow model and the lattice definition of
its magnetic monopoles in Section~\ref{sect:1Higgs}.  In
Section~\ref{sec:expectations}, we use semi-classical results to
motivate our numerical results. We present details of the Monte Carlo
simulations carried out in Section~\ref{sec:numerics}, and the results
obtained in Section~\ref{sec:results}. Finally we discuss our findings
in Section~\ref{sec:conclusions}.

\section{The Georgi--Glashow model}
\label{sect:1Higgs}

In the continuum, the three--dimensional Georgi--Glashow model 
is defined by the Lagrangian
\begin{equation}
{\cal L} = \frac{1}{2} \Tr \left( F_{ij} F_{ij} \right) + 
\Tr [D_i, \Phi] [D_i,\Phi] + m^2 \Tr \Phi^2 +
\lambda \left( \Tr \Phi^2 \right)^2,
\end{equation}
where $\Phi$ is in the adjoint representation of the SU(2) gauge
group, $D_i=\partial_i+ig_3A_i$ and $F_{ij}=(ig_3)^{-1}[D_i,D_j]$.
The partition function of the theory is formally defined as the path
integral
\begin{equation}
{\cal Z}=\int D\Phi DA_i\exp\left(-\int d^3x {\cal L}\right).
\end{equation}
This can be interpreted as a three--dimensional Euclidean quantum
field theory, or as a classical statistical field theory with the
Hamiltonian $\beta H=\int d^3x {\cal L}$.

The coupling constant, $g_3^2$, has the dimensions of mass, and we can
write the parameters of the theory in terms of dimensionless ratios
with the coupling constant
\begin{equation}
\label{equ:xdef}
x = \frac{\lambda}{g_3^2}
\end{equation}
and
\begin{equation}
\label{equ:ydef}
y = \frac{m^2(g_3^2)}{g_3^4}.
\end{equation}
The notation here reflects the fact that the theory is
super--renormalisable (in three dimensions), and thus only the scalar
mass needs a renormalisation counterterm. Even this is only necessary
up to the two loop level, and its value is known both in the
$\overline{\mbox{MS}}$ scheme
\cite{farakos94}
and in lattice regularisation
\cite{Laine:1995ag,laine97}.
In Eq.~(\ref{equ:ydef}), $m^2(g_3^2)$ is the $\overline{\mbox{MS}}$
renormalised mass with renormalisation scale $\mu=g_3^2$.

To study this model in a fully non--perturbative manner, we formulate
the theory in a way that allows numerical solution by Monte Carlo
simulation on a cubic, Euclidean lattice consisting of $L^3$ sites,
labelled by a triplet of integers $\vec{x}=(x,y,z)$.  The action is
given by $S = \sum_{\vec{x} } {\cal L}(\vec{x})$, with the Lagrangian
\begin{eqnarray}
{\cal L}(\vec{x}) & = &
\beta\sum_{i<j}\left[1-\frac{1}{2}
\tr U_{ij}(\vec{x}) 
\right]\nonumber 
\\
&&+\sum_{i}\left\{2a \left[\tr\Phi^2(\vec{x}) - 
\tr \Phi(\vec{x}) U_i(\vec{x}) \Phi(\vec{x}+\hat{\imath}) 
U_i^\dagger(\vec{x}) \right]\right.
\nonumber
\\
& &\left. + m^2 a^3 \tr \Phi^2(\vec{x}) + a^3 \lambda
\left[\tr \Phi^2(\vec{x})\right]^2\right\},
\label{equ:lagr}
\end{eqnarray}
where $m^2$ is the bare lattice mass parameter and $\beta=4/(ag_3^2)$
is the conventional notation for the bare lattice gauge coupling.

We shall treat this lattice theory as an approximation to the
continuum one, and therefore we parameterise the theory in terms of
the renormalised continuum couplings defined in Eqs.~(\ref{equ:xdef})
and (\ref{equ:ydef}).  We are able to do this, because the
relationships between the lattice and continuum couplings are
known~\cite{Laine:1995ag,laine97}, but we shall postpone discussion of
them until Section~\ref{sec:numerics}. We shall also express all
quantities in continuum units.

\subsection{Magnetic monopoles}

It is very well known that in the continuum, the field equations have
topologically non-trivial solutions, 't~Hooft--Polyakov monopoles
\cite{'tHooft:1974qc,Polyakov:1974ek}.
They can be characterised by a non-zero winding number of the Higgs
field at the spatial infinity,
\begin{equation}
N_W=\frac{1}{16\pi i}\int d^2S_k \epsilon_{ijk}\tr\ph\left(\partial_i\ph\right)
\left(\partial_j\ph\right)\in \mathbb{Z},
\end{equation}
where $\ph=\Phi(\Phi^2)^{-1/2}$.  Although $N_W$ itself is gauge
invariant, the integrand is not, and therefore it does not have a
direct physical interpretation.  It can be easily seen, however, that
$N_W$ actually corresponds to the magnetic charge associated with the
residual U(1) gauge invariance.

To see this, let us define the magnetic field as~\cite{'tHooft:1974qc}
\begin{equation}
\label{equ:U1field}
{\cal B}_{i}=\frac{1}{2}\epsilon_{ijk}\left[\Tr\ph F_{jk}
+\frac{1}{2ig}\Tr\ph (D_j\ph)(D_k\ph)\right].
\end{equation}
This is a gauge invariant quantity, and agrees with
$\vec{\nabla}\times\vec{A}_3$ in the unitary gauge $\ph=\sigma_3$.
Therefore it is indeed the magnetic field associated with the residual
U(1) symmetry.  The corresponding magnetic charge density,
$\rho_M=\vec{\nabla}\cdot\vec{\cal B}$, has the following properties:
First, because $\rho_M$ is given by a total derivative, the charge
inside a given volume can be expressed as a surface
integral. Therefore any local deformation of the fields inside the
volume cannot change the charge inside the volume.  Second, the
magnetic charge inside a given volume is, in fact,
\begin{equation}
Q_M=\int d^3x \rho_M=\frac{4\pi}{g}N_W,
\end{equation}
and is therefore quantised in units of $4\pi/g$.  These two properties
imply that the only way the charge inside a volume can be changed is
by moving a magnetic monopole in or out of the volume.  In other
words, the magnetic charges are topologically stable.

What is less well known is that these same properties are also true
for the lattice theory. We can define the analogue of
Eq.~(\ref{equ:U1field}) as
\begin{equation}
\hat{B}_i=\epsilon_{ijk}\alpha_{jk}.
\end{equation}
Here $\alpha_{jk}$ is the lattice U(1) field strength tensor,
\begin{equation}
\alpha_{ij}=\tr\Pi_+(\x)U_i(\x)\Pi_+(\xpi)U_j(\xpi)
\Pi_+(\xpij)U^\dagger_i(\xpj)
\Pi_+(\xpj)U^\dagger_j(\x),
\end{equation}
and $\Pi_+=\frac{1}{2}(1+\ph)$.  In the continuum limit, $\hat{B}_i$
approaches $a^2{\cal B}_i$. If we define the magnetic charge inside a
lattice cell as
\begin{equation}
\hat{\rho}_M(\x)=\sum_i\left[\hat{B}_i(\xpi)-\hat{B}_i(\x)\right],
\end{equation}
it satisfies the same conditions that guarantee in the continuum the
topological stability of magnetic monopoles: the charge is quantised
and can be written as a surface integral.  
These are the same properties that ensure the stability of monopoles
in the continuum, and thereby magnetic
monopoles are well-defined and absolutely stable 
objects even in a discrete
lattice theory, unlike the instantons of the four-dimensional
Yang-Mills theory.
Because of the quantisation and stability of magnetic charge, it makes
sense to consider `microcanonical' partition functions $Z_{Q_M}$
which are restricted to configurations with a given magnetic charge
$Q_M$.  The full, `canonical' partition function is then simply
\begin{equation}
{\cal Z}=\prod_{Q_M=-\infty}^{\infty}Z_{Q_M}.
\end{equation}
We define the free energy of a given topological sector by
\begin{equation}
F_{Q_M}=-\ln Z_{Q_M},
\end{equation}
and the free energy of a monopole as the free energy difference of
sectors $Q_M=1$ and $0$,
\begin{equation}
\label{equ:fediff}
\Delta F=F_{1}-F_0.
\end{equation}
Semiclassically, $Z_{Q_M}=\exp[-S(Q_M)]$, where $S(Q_M)$ is the action
of the monopole solution with charge $Q_M$.  $S(Q_1)$ can also be
interpreted as the mass of a monopole, and with a slight abuse of
language we can generalise into the fully non-perturbative case by
defining the monopole `mass' $M$ by
\begin{equation}
M=g_3^2\Delta F.
\end{equation}
The semiclassical picture would predict that monopoles are massive in
the broken phase and massless in the symmetric phase.  If this were
true, the mass would serve as an order parameter for the phase
transition.

\subsection{Boundary conditions}

We measure the monopole free energy following the method of Ref.~%
\cite{Davis:2000kv},
which for convenience we briefly review in this Section.

Our strategy is to work on a finite sized system, and impose boundary
conditions that force the total magnetic charge of the lattice to be
either odd or even, whilst preserving the translation invariance of
the system.  This is important because translation invariance
guarantees the absence of boundary effects.

Gauss's law rules out periodic boundary conditions, as the total
charge is constrained to be zero. However,
translation invariance is preserved by any boundary conditions that
are periodic up to symmetries of the Lagrangian, and in general they
allow a non-zero magnetic charge.  For instance,
`C--periodic boundary conditions'
\cite{Kronfeld:1991qu}
\begin{equation}
\label{equ:ccbc}
\Phi(n+L\jh) = - \sigma_2 \Phi(n) \sigma_2 = \Phi^*(n), \qquad
U_k(n+L\jh) = \sigma_2 U_k(n) \sigma_2 = U_k^*(n).
\end{equation}
are such that the net magnetic charge can be non-zero, but it is
constrained to be even~\cite{Davis:2000kv}.  We shall refer to
calculations using such boundary conditions with a subscript `0'.

Similarly, if the fields are constrained to behave as 
\begin{equation}
\Phi(n + L\jh) = - \sigma_j \Phi(n) \sigma_j, \qquad
U_k(n + L\jh) = \sigma_j U_k(n) \sigma_j.
\label{equ:bc}
\end{equation}
on moving around the lattice, the net magnetic charge is odd. We term
these `twisted (C--periodic) boundary conditions', and denote results
so obtained by a subscript `1'. It is easy to see that both sets of
boundary conditions are symmetries of the lattice Lagrangian.

By a gauge transformation, the twisted boundary conditions may be
rewritten as (untwisted) C--periodic boundary conditions everywhere
save at the edges of the lattice, where
\begin{eqnarray}
U_3(x,L,L-1) &=& -U_3{}^*(x,0,L-1)
,\nonumber\\
U_1(L-1,L,z) &=& -U_1{}^*(L-1,0,z)
,\nonumber\\
U_1(L-1,y,L) &=& -U_1{}^*(L-1,y,0)
.
\end{eqnarray}
By a suitable redefinition of the fields
\begin{eqnarray}
\label{equ:redef}
U_3(x,N,N-1) &\rightarrow&-U_3(x,N,N-1)
,\nonumber\\
U_1(N-1,N,z) &\rightarrow&-U_1(N-1,N,z)
,\nonumber\\
U_1(N-1,y,N) &\rightarrow&-U_1(N-1,y,N)
,
\end{eqnarray}
we can express the twisted boundary conditions as a theory with
C--periodic boundary conditions everywhere, but with an additional
term in the action that depends solely on the gauge fields:
\begin{equation}
\label{equ:freeen1}
Z_1 = \int DU_i D\Phi \exp \left(-S - \Delta S \right),
\end{equation}
where the change in the action is
\begin{eqnarray}
\label{equ:monodeltaS}
\Delta S&=&
\beta \left[
\sum_{x=0}^{L-1} \tr U_{23}(x,y_0,z_0)
\nonumber \right.
\\
& & \left.
+
\sum_{y=0}^{L-1}\tr U_{13}(x_0,y,z_0)
+
\sum_{z=0}^{L-1}\tr U_{12}(x_0,y_0,z)
\right].
\end{eqnarray}
We emphasise that, because Eq.~(\ref{equ:freeen1}) is equivalent to
Eq.~(\ref{equ:bc}) with the translation invariant boundary conditions,
the choice of coordinates $(x_0,y_0,z_0)$ does not affect any
observable, and in particular, it does not fix the location of the
monopole on the lattice.

In physical terms, $\Delta S$ gives a negative gauge coupling to three
orthogonal stacks of plaquettes which are pierced by three mutually
intersecting lines on the lattice. These lines are known in the
literature as 't~Hooft lines~\cite{'tHooft:1978hy}. A
single, open 't~Hooft line creates a pair of Dirac monopoles, and has
been used to measure their interaction potential in
Refs.~\cite{Kovacs:2000sy,Hoelbling:2001su,hart00d,deForcrand:2001fi,%
Chernodub:2001nz}.  It should be noted, however, that Dirac monopoles
are rather different from 't~Hooft-Polyakov monopoles. They have only
half the magnetic charge of the latter, and are singular,
non-dynamical objects.  In our case, the 't~Hooft lines are closed by
the boundary conditions, and therefore they do not create any
singularities, but a non-singular 't~Hooft--Polyakov monopole.

The free energy $\Delta F$ (or the `mass')
of a monopole is defined by analogy with
Eq.~(\ref{equ:fediff}) as
\begin{equation}
\label{equ:unprog}
\Delta F = F_1 - F_0 
\equiv - \ln \left( \frac{Z_1}{Z_0} \right)
= -\ln \langle \exp (-\Delta S) \rangle.
\end{equation}
In the main, however, we shall study the derivative of $\Delta F$ with
respect to the scalar mass parameter, $y$,
\begin{equation}
\label{equ:Mderiv}
\frac{1}{g_3} \frac{\partial \Delta F}{\partial y} = g_3^6 V 
\left( \frac{\langle \tr \Phi^2 \rangle_1}{g_3^2} -
\frac{\langle \tr \Phi^2 \rangle_0}{g_3^2} \right),
\end{equation}
where $V$ is the volume of the system. We know that for sufficiently
large $y$ in the symmetric phase the free energy of the monopole will
go to zero (at least in the large volume limit). If we see the
derivative becoming zero, the free energy is at most a constant. In
Section~\ref{sec:numerics} we also measure the free energy at a point
in the symmetric phase and find it to be consistent with zero. If the
derivative is zero all over the symmetric phase, it is reasonable to
assume, then, that the free energy itself is becoming zero.

\section{Semiclassical expectations}
\label{sec:expectations}

We now turn our attention to the semi--classical predictions for the
't Hooft--Polyakov monopole (see, for instance,
\cite{polyakov87}),
to which we would like to compare our results from the fully quantised
theory.

In the broken phase of the theory the scalar field gains a vacuum
expectation value (VEV)
\begin{equation}
\vev = \sqrt{\frac{-y}{2x}}.
\end{equation}
The semi--classical solution of unit winding number is the 't
Hooft--Polyakov monopole
\cite{'tHooft:1974qc,Polyakov:1974ek},
associated with an isolated zero of the scalar field. Away from this,
the scalar field decays towards its VEV, with a characteristic length
scale
\begin{equation}
\xi_s g_3^2 = \left( \sqrt{-y} \right)^{-1}.
\end{equation}
The gauge field simultaneously decays from being SU(2) to being
asymptotically U(1) with a length scale
\begin{equation}
\xi_g g_3^2 = \left( \vev \right)^{-1}.
\end{equation}
We thus have a picture where asymptotically the gauge fields are
Abelian, save within some extended core whose size is defined by the
above length scales where the gauge fields `unwind' into the full
SU(2) gauge manifold. We shall find that this scenario remains at
least qualitatively valid when quantum corrections are introduced.

The mass of this object is, semiclassically,
\begin{equation}
\frac{M}{g_3^2} = 4 \pi \vev f(x)
\label{eqn_sc_mass}
\end{equation}
where $f(x)$ is the 't Hooft function. To satisfy the Bogomolny
lower bound on the mass, $f(0) = 1$. Also, it is known numerically
(see, for example,
\cite{goddard78})
that for small $x$, $f(x) \simeq 1 + x$.

The derivative of this mass, as in Eq.~(\ref{equ:Mderiv}), is
\begin{equation}
\frac{p_0}{g_3^2} \equiv \frac{\partial}{\partial y} \frac{M}{g_3^2} = 
-\frac{2 \pi}{\sqrt{-2xy}}.
\label{equ:sgl-deriv}
\end{equation}

Assuming that the monopoles are point-like and non-interacting, we can
roughly estimate their density to be
\cite{Polyakov:1977fu}
\begin{equation}
\frac{\nu_0}{g_3^6} = \left( \vev \right)^{\frac{7}{2}} 
\exp\left( -\frac{M}{g_3^2} \right),
\label{equ:density}
\end{equation}
which is suppressed by the exponential of the mass.
We may also define a mean separation $D$ of the monopoles as 
\begin{equation}
\frac{1}{\left( D g_3^2 \right)^3} \equiv \frac{\nu_0}{g_3^6}.
\end{equation}
When $M\gg g_3^2$, there is a hierarchy between $D$ and the
fundamental length scales $\xi_s$ and $\xi_g$, and therefore
the above assumption of point-like monopoles is valid.

This is the semiclassical picture for infinite volume. What we are
interested in is what happens in the quantised theory of finite
volume, and the interplay of the system size, $aL$ with the scales
above. Particularly, we wish to know the fate of the monopole mass on
large length scales.

\subsection{Small volumes}
\label{sect:small}
Let us first briefly discuss what happens when the volume of the
system is comparable to, or smaller than, the length scales discussed
before.  The core size of a monopole is given by the correlation
length $\xi$ (we assume for sake of argument that $\xi_g$ and $\xi_s$
are comparable), and therefore if $aL \lesssim \xi$, there is no room
for a monopole in the system. If the system is forced by twisted
boundary conditions to contain one monopole, its core will fill the
lattice and the whole system will be in the confining phase. On the
other hand, the untwisted system is in the Higgs phase. The free
energy densities of these two phases differ by a certain non-zero
amount $\Delta f$, which is essentially the latent heat, and as this
is the case in the whole volume, we have
\begin{equation}
\Delta F\approx L^3\Delta f.
\end{equation}
Thus, we can conclude that when $aL\ll\xi$, $\Delta F$ should scale as
the volume of the system.

When the volume is increased, $\xi \lesssim aL \lesssim 2.5 \xi$, the
fields start to approach the U(1) of the Higgs vacuum far from the
monopole core of the twisted system.  Nonetheless, the core will be
affected by the boundary conditions, and in general a restriction in
the core size by the boundary will lead to an increase in the
(absolute value of the) free energy and its derivatives. As a rough
estimate, if the total non--Abelian flux inside the monopole core is
roughly constant, then the flux density will vary as the inverse of
the volume. The total energy of the system would then vary as
$L^{-3}$. (The figure $2.5 \xi$ is a rough limit derived from our
results.)

\subsection{Intermediate volumes}

Let us then consider a system that is large enough to comfortably
accommodate one monopole, but is so small that the fluctuations are
not likely to create isolated monopoles (or, more accurately, well
separated monopole--antimonopole pairs).  This is the case when $\xi
\ll aL \ll D$. That is, the entropy--action balance is dominated by
the action cost, which limits us to the minimum number of monopoles
(and antimonopoles) required to satisfy the boundary conditions.

We expect the free energy difference, Eq.~(\ref{equ:unprog}), to be
that between a system of one monopole and an uncharged box.  Because a
monopole is a localised object, the regions far from the monopole core
are unaware of the twist in the boundary conditions. $\Delta F$ only
gets a contribution from the monopole core and is therefore
independent of the volume.  In this case, the identification of
$\Delta F$ with the monopole mass makes sense, and a comparison
between the measured values and the semiclassical formul\ae~above
yields information on the radiative corrections to the semiclassical
monopole.

\subsection{Large volumes}

As the volume is increased such that $aL \gg D$, the entropy gain in
introducing well separated monopole--antimonopole pairs into the
vacuum outweighs the action cost and the mean density of topological
objects is no longer expected to be the minimum commensurate with the
boundary conditions. The free energy required to introduce an extra
monopole into the system is now less than the mass of the single
monopole, as we demonstrate with a simple model.

\subsubsection{The dilute monopole gas}

Following Ref.~\cite{hart00c}, where a similar effect was discussed in
the case of vortices in (2+1)--dimensions, we assume that the density
of monopoles is low enough, so that the
probability of finding one in any sub-volume of space is independent
of whether there are monopoles present elsewhere in the system.  In
other words, the monopoles are assumed to be point-like or that
overlap of the cores is of vanishing measure.  As discussed above,
this dilute monopole gas approximation is believed to be valid deep in
the broken phase.

The probability of finding $n$ monopoles or antimonopoles (we do not
distinguish) in a volume, $V$, follows Poissonian statistics
\begin{equation}
p(n;V) = \frac{1}{\cal N} \frac{1}{n!} 
\left( \nu_0 V \right)^{n}.
\end{equation}
We apply this to the volume of the whole lattice
$V=(aL)^3=g_3^{-6}(4L/\beta)^3$.  We find different normalisation
factors for twisted ($n \in {\rm odd}$, and ${\cal N}_1 = \sinh \left(
\nu_0 V \right)$) and untwisted ($n \in {\rm even}$, and ${\cal N}_0 =
\cosh \left( \nu_0 V \right)$) boundary conditions.

The free energy of the system (or its derivative) is extensive and the
sum of the free energy of the components for a dilute gas, and
considering the entire system we obtain:
\begin{eqnarray}
\frac{\partial}{\partial y} \frac{\Delta F}{g_3^2} 
& = &
\frac{1}{{\cal N}_1} \sum_{n \in {\rm odd}} 
n \left( \frac{p_0}{g_3^2} \right) p(n;V)
\nonumber\\&& - 
\frac{1}{{\cal N}_0} \sum_{n \in {\rm even}} 
n \left( \frac{p_0}{g_3^2} \right) p(n;V)
\nonumber
\\
& = &
4 \left( \frac{p_0}{g_3^2} \right) \nu_0 V 
\frac{e^{-2 \nu_0 V}}{1-e^{-4 \nu_0 V}}
\label{equ:multi-deriv}
\end{eqnarray}
where $p_0$ was defined in Eq.~(\ref{equ:sgl-deriv}).
Eq.~(\ref{equ:multi-deriv}) gives the desired plateau for intermediate
$V$, but then decays to zero as $V \to \infty$, beginning once $V
\gtrsim V_c$, such that $\nu_0 V_c = 1$. Note that since $\nu_0 V$ is
simply the typical number of monopoles and antimonopoles created by
fluctuations, this result shows that the monopole free energy decays
as soon as the fluctuations can create isolated monopoles.

Crucial in the above calculation is the assumption that the monopoles
are non-interact\-ing. Although the monopoles at least semiclassically
have a long range Coulomb interaction, we believe this approximation
is justified, because the interaction is non-confining.  Nevertheless,
it is only an approximation, and therefore it must be tested in
numerical simulations, as we do in Sect.~\ref{sec:results}.

We can also see that the above argument would break down if we tried
to apply it to the four--dimensional case, where the monopoles are
world lines rather than point-like objects. In a Euclidean theory, the
action of the monopole world line would be proportional to its length,
and therefore $\nu_0$ would vanish exponentially when the limit of
infinite time dimension is taken. The same happens for vortices in the
three--dimensional U(1) theory~\cite{Kajantie:1999zn}. In future work
we aim to verify that this is also true for vortices in a non--Abelian
theory.  On the other hand, if one of the three dimensions is compact
as in the (2+1)--dimensional case at a non-zero temperature, $\nu_0$
is finite, and again the vortex free energy vanishes in the infinite volume
limit~\cite{hart00c}.

\subsubsection{Confinement}

The prediction of the dilute monopole gas approximation that the
monopole free energy vanishes in both phases in the infinite volume
limit is compatible with the properties the phase diagram of the
theory is believed to have. Vanishing free energy means that the
monopoles condense, and according to the dual superconductor
picture~\cite{Mandelstam:1976pi}, this gives rise to confinement. 

Indeed, it is known semiclassically that the non-zero monopole density
gives the photon a non-zero mass even in the Higgs
phase~\cite{Polyakov:1977fu}, and this leads to confinement. Thus it
is natural to assume that the Higgs phase is analytically connected to
the confining phase~\cite{Nadkarni:1988pb,Hart:1997ac}.  Again, this
can only be true if the monopole free energy vanishes in the Higgs
phase, because otherwise it would act as an order parameter signalling
a transition from the Higgs to the confining phase.

Previous studies \cite{Hart:1997ac} have supported the idea of a smooth
crossover between the phases, but as they only concentrated on local
quantities, they cannot be regarded as proofs. For instance, in the
three--dimensional Abelian Higgs model, the phase transition can only
be seen in practice by measuring non-local observables such as
the vortex tension or the photon mass~\cite{Kajantie:1999zn}.
In the present case, the predicted non-zero photon mass has
not been observed in simulations~\cite{Hart:1997ac,Kajantie:1997tt}.
It is clear from the results presented here that the
reason for this lies in the very large volumes required.

\section{Lattice Monte Carlo simulations}
\label{sec:numerics}

We simulate the Georgi--Glashow model on the lattice via Monte Carlo
importance sampling of the partition functions for both the
C--periodic and twisted C--periodic boundary conditions.
Updates to the lattice were performed as compound sweeps consisting of
one heatbath update to the gauge and scalar fields, followed by two
over-relaxation steps to each. Measurements were made once per
compound sweep.

Statistical errors were estimated by jack-knife analysis, dividing the
data sets into ten bins. For most lattices, the bin size was much
longer than the autocorrelation time of the observables, making them
independent. This could be seen in an approximate decrease in the
statistical errors as $1/\sqrt{N}$ as the number of measurements, $N$,
was increased. The only lattice on which this was not readily apparent
was the $\beta=4.5$, $L=46$, where the errors did not show such a
reduction. This may indicate that, despite considerable computational
effort, the ensemble size is still such that the autocorrelation time
was comparable to the bin size. Error estimates for this ensemble
should thus be treated as lower bounds.

\begin{figure}
\epsfig{file=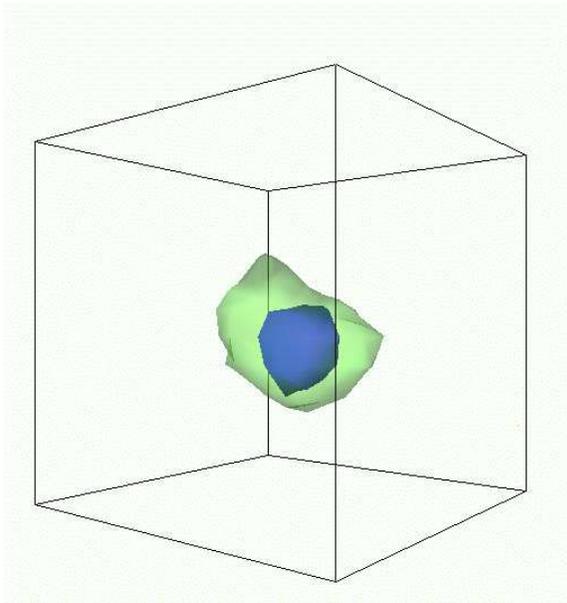,width=7.5cm}
\caption{
\label{fig:monopole}
Isosurfaces ${\rm Tr}\,\Phi^2=3.89$ (green) and
$\sum_{i<j}(1-{\rm Tr} U_{ij}/2)=0.18$
(blue) in a
typical field configuration at $x=0.05$, $y=0.45$ and $\beta=18$.
In order to reduce noise, the configuration was averaged over 50
subsequent Monte Carlo sweeps.}
\end{figure}

To illustrate that the twisted boundary conditions
(\ref{equ:bc}) indeed generate a mono\-pole, we show in
Fig.~\ref{fig:monopole} the isosurfaces of
${\rm Tr}\,\Phi^2$ and the gauge action density $\sum_{i<j}(1-{\rm Tr}
U_{ij}/2)$ in a typical field
configuration at $x=0.05$, $y=0.45$ and $\beta=18$.
The gauge action peaks and ${\rm Tr}\,\Phi^2$ dips around the same point,
exactly as is expected to happen near the mono\-pole core.
Because of thermal fluctuations, the isosurfaces are not spherical.

\subsection{Observables}

We measure the free energy and its derivative with respect to the
scalar mass. The former is done via Eq.~(\ref{equ:unprog}). In
practice this does not work; the importance sampling of the theory with
untwisted boundary conditions has very small overlap with that of the
twisted partition function. This leads to strong sign fluctuations in
$\Delta S$ which leads to a poor convergence of its average through
Monte Carlo simulation.

Instead, as in
Refs.~\cite{Kajantie:1999zn,Kovacs:2000sy,Hoelbling:2001su}, we can
introduce a set of ensembles defined by a real parameter, $\varepsilon
\in [0,1]$:
\begin{equation}
Z_\varepsilon \equiv 
\int DU_i D\Phi \exp \left( -S - \varepsilon \Delta S \right),
\label{equ:freeeneps}
\end{equation}
where $\varepsilon = 0$ is the untwisted case, and $\varepsilon = 1$
represents twisted boundary conditions. We then write
\begin{equation}
\label{equ:monomass}
\Delta F = \int_0^1 d\varepsilon 
\frac{\partial F_\varepsilon}{\partial \varepsilon}
= \int_0^1 d\varepsilon \langle \Delta S \rangle_\varepsilon,
\end{equation}
where the subscript $\varepsilon$ indicates that the expectation value
must be measured using Eq.~(\ref{equ:freeeneps}).  This gives us the
absolute value of $\Delta F$, but with the cost that we have to
measure expectation values at non-physical values of $\varepsilon$. We
call this the `method of progressive twisting'.  Calculations of the
free energy by progressive twisting typically used 10,000 to 20,000
measurements for each of 37 values of the twisting parameter,
$\varepsilon$, which are then numerically integrated. (For an
alternative approach, see Ref.~\cite{deForcrand:2001fi}.)

Alternatively, the derivative of the free energy,
Eq.~(\ref{equ:Mderiv}), may be measured directly, which avoids the
reweighting problem. We are, however, calculating an intensive
quantity as the difference of two approximately extensive
numbers. Maintaining a constant error on the former demands increasing
accuracy in the latter for increasing volume. Even allowing for
self--averaging and the good scaling properties of the simulation
algorithm, maintaining comparable precision in the free energy
derivative requires CPU time rising as $L^6$. This limits the results
of this study to $L \le 46$. Calculations of the derivative with
respect to the scalar mass used between 200,000 and 500,000
measurements for each of the boundary condition choices.

\subsection{Lattice parameters}

\begin{table}
\begin{tabular}{|l|l|l|l|l|}
\hline \hline
\multicolumn{1}{|c|}{$x$} & 
\multicolumn{1}{c|}{$y$} & 
\multicolumn{1}{c|}{$\beta$} & 
\multicolumn{1}{c|}{$a g_3^2$} & 
\multicolumn{1}{c|}{$L$} \\
\hline
0.35 & $-10$, $-3$, $-1$, $-0.5$, 1, 10 & $18.0$ & $0.222$ & 16 \\
\hline \hline
\end{tabular}
\caption{\label{tab:prog-params}
Lattices used to study the
monopole free energy by the method of progressive twisting.}
\end{table}

\begin{table}
\begin{tabular}{|l|l|l|l|l|}
\hline \hline
\multicolumn{1}{|c|}{$x$} & 
\multicolumn{1}{c|}{$y$} & 
\multicolumn{1}{c|}{$\beta$} & 
\multicolumn{1}{c|}{$a g_3^2$} & 
\multicolumn{1}{c|}{$L$} \\
\hline
0.35 & $-0.124$ & 
4.5 & 0.889 &
4, 6, 8, 10, 12, 14, 16, 18, 20, \\
&&&&
22, 24, 28, 32, 36, 40, 46 \\
&&
6.0 & 0.667 &
4, 6, 8, 10 \\
&&
9.0 & 0.444 &
4, 6, 8, 10, 12, 14, 16 \\
&&
12.0 & 0.333 &
4, 6, 8, 10, 12, 14, 16, 20 \\
&&
18.0 & 0.222 &
4, 6, 8, 10, 12, 14, 16, 20 \\
\hline \hline
\end{tabular}
\caption{\label{tab:deriv-params}
Lattices used to study the system
size dependence of the derivative of the monopole free energy.}
\end{table}

The physical and lattice parameter values used are
listed for reference in Tables~\ref{tab:prog-params}
and~\ref{tab:deriv-params}.  

In this section we discuss simulations of the SU(2) Georgi--Glashow
model in three Euclidean dimensions. The action for the theory has
been given in Eq.~(\ref{equ:lagr}).  In addition to the parameters
that define our theory in the continuum limit, $x$ and $y$, there are
two additional complications in the lattice theory, being the lattice
spacing, $a g_3^2$, and the volume, $(a g_3^2 L)^3$, of the cubic
lattice on which we perform the simulations.

Detailed investigations of finite volume effects and scaling of
correlation lengths have been performed for the $d=2+1$ pure gauge
SU(2) and Georgi--Glashow field theories in
\cite{teper98,hart99}.
Here we summarise the findings briefly for the benefit of
non--specialist readers. 

The lattice calculations yield dimensionless results, which may be
interpreted as being the physical result multiplied by the lattice
spacing raised to their na\"{\i}ve dimensions, and which we denote via
a circumflex accent. We remove the dependence on the unknown lattice
spacing by multiplying the result with the appropriate power of
$\beta=4/(ag_3^2)$, and therefore it is natural to express the results
in terms of powers of $g_3$, which has the dimensions of
(mass)$^{1/2}$.  For sufficiently fine lattices, the agreement with
the continuum limit will be within the statistical errors of the
lattice data, but on coarser lattices there may in principle be
deviations.  The results in Refs.~\cite{teper98,hart99} are indicative
of the continuum limit for $\beta \gtrsim 4.5$, which includes
relatively coarse lattices at the lower end of this range (as we
discuss later).

The lattice theory in Eq.~(\ref{equ:lagr}) is parameterised by three
couplings $(m^2,\lambda,\beta)$.  In order to vary the lattice
spacing, we wish to change $\beta$ whilst maintaining the same
continuum theory [\ie\ $(x,y)$].  This is commonly referred to as
moving along `lines of constant physics'.  These trajectories have
been calculated~\cite{Laine:1995ag,laine97} in the limit
$\beta\rightarrow\infty$, and they are believed to be valid for
lattices finer than $\beta \simeq 4.5 - 5.0$:
\begin{eqnarray}
\beta &=& \frac{4}{ag_3^2},\nonumber\\
\lambda &=& x g_3^2,\nonumber\\
\frac{m^2}{g_3^4} &\approx& y-\left(4+5x\right)\frac{3.1759}{4\pi ag_3^2}
\nonumber\\
&&
-\frac{1}{16\pi^2}\Biggl[
\left( 20x-10x^2 \right)
\left(\ln\frac{6}{ag_3^2}+0.09\right)
\nonumber\\
&&
~~~~~~~~~~+11.6x+8.7
\Biggr].
\end{eqnarray}
Again, we address the range of applicability in a later section.  We
are primarily interested in testing the idea that the 't
Hooft--Polyakov monopoles condense. The measurement of this is a fine
balance. Whilst monopoles are topologically stable even if their core
is smaller than the lattice spacing, it should be much larger than
that to ensure they resemble the semiclassical 't~Hooft--Polyakov
solution.  Experience indicates that the correlation lengths of the
gauge and scalar field should be at least 2 or 3 lattice spacings.
Simultaneously, in order to see the screening of the free energy that
signals the formation of the plasma, we require lattices that are
(much) larger than the mean separation of the monopoles, such that it
is possible for screening of magnetic charge to occur. Given that
these two scales may be widely separated, it is not at all clear that
we will be able to achieve the balance using a lattice size, $L$,
which can be realistically simulated on the resources available.

We can use the known, semi-classical description of the monopoles
\cite{Polyakov:1977fu}
to estimate the parameters needed for the lattice. Such estimates are,
of course, only expected to be accurate up to numerical factors which
may be important here. Nonetheless, we may hope the results are
indicative at least, and the exercise gives some insight into the
possible screening mechanism.

The monopole density (\ref{equ:density}) has a maximum value of
just under $0.000345$. Screening will become apparent when the
physical volume, $g_3^6 V \equiv \left( 4L/\beta \right)^3$, is such
that $\nu_0 V_c = 1$; this yields
\begin{eqnarray}
\frac{4L_c}{\beta} 
& \gtrsim &
\left( \left. \frac{\nu_0}{g_3^6} \right|_{\rm max} 
\right)^{-\frac{1}{3}}, \\
L_c
& \gtrsim &
3.56 \beta.
\end{eqnarray}
If a conservative value of $L_c = 16$ is chosen to allow for possible
suppression of the monopole density, this indicates that the gauge
coupling is restricted to be $\beta\le 4.5$.  Our primary interest is
in observing the monopole screening, so it is not strictly necessary
that the perturbative lines of constant physics still hold on our
lattices. We would like to maintain some contact with continuum
physics, however, and thus go no lower than $\beta = 4.5$.

Using Eq.~(\ref{equ:density}), the maximum monopole density is reached
for $v/g_3 = 0.421$. We are most interested in the fate of the
monopole mass in the region of the phase diagram where there is a
crossover between the two phases. For this reason we select $x=0.35$,
and thus $y=-0.124$.  At this parameter set, the gauge correlation
length is, in units of the lattice spacing,
\begin{equation}
\hat{\xi}_g = \frac{\beta}{4} \left( \vev \right)^{-1} = 2.67,
\end{equation}
and the scalar field correlation length is
\begin{equation}
\hat{\xi}_s = \frac{\beta}{4} \left( \sqrt{-y} \right)^{-1} = 3.20,
\end{equation}
which are both suitably larger than the lattice grid size. Finer
lattices were used to resolve better the small volume behaviour.

\section{Results}
\label{sec:results}

\subsection{Small volumes}

\begin{figure}
\epsfig{file=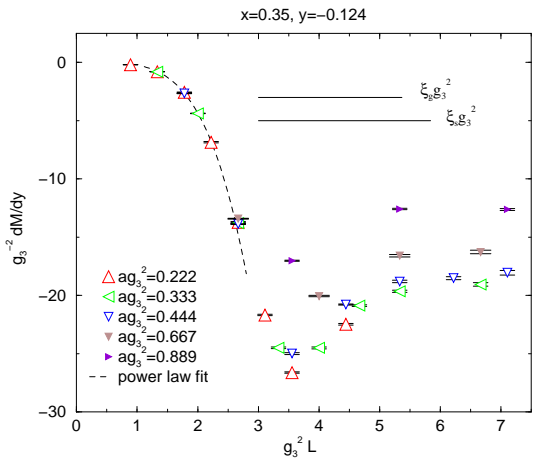,width=7cm}
\caption{\label{fig:small-L}
The derivative of the monopole
free energy as a function of (small) lattice size at $x = 0.35, y =
-0.124$.}
\end{figure}

As discussed in Sect.~\ref{sect:small} the free energy $\Delta F$ is
expected to be proportional to the volume of the system when
$L\lesssim \xi$. We studied this in our simulations by measuring its
$y$-derivative with couplings $x=0.35$ and $y=-0.124$.  Obviously,
this should behave in the same way as the free energy difference
itself.  The results from lattices of different sizes and different
lattice spacings are plotted in Fig.~\ref{fig:small-L} as functions of
the physical lattice size $ag_3^2L$. At small $L$ the data show very
little scaling violation. This suggests that we are not seeing a
physically interesting effect here and supports the idea that the
behaviour with $L$ has a simple origin. We show a fit of the form
$-d_0L^{d_1}$, where $d_{\{0,1\} }$ are free parameters. Whilst the
power law fits well by eye, the precise nature of the data makes the
fits all quite poor ($\chi^2/{\rm dof} \gtrsim 5$). The fit shown is
to the $\beta=18.0$ data only, and gives $d_1 = 4.2 \, (5)$. Whilst
not precisely 3, this gives qualitative support to our simple picture.

\begin{figure}
\epsfig{file=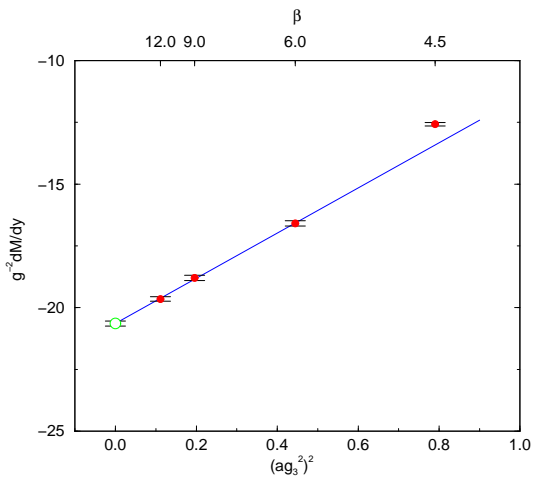,width=7cm}
\caption{\label{fig:plat_scal}
Continuum limit of the derivative of the free energy at
$x=0.35$, $y=-0.124$ for fixed physical volume, $g_3^2L=5.3$.}
\end{figure}

Beyond $g_3^2 L \simeq 3.5$ we see different behaviour. The derivative
now decreases towards a plateau on intermediate scales. Whilst this
decay may be a power law, we find the data insufficient to support a
precise fit. The value of the plateau does show evidence of a
discretisation effect. We may attempt to quantify this through a
continuum extrapolation of the data at $g_3^2L = 5.3$, admittedly
still in the transient region, but where we have results for four
couplings. We show the data in Fig.~\ref{fig:plat_scal}, along with a
fit assuming only a leading order correction to scaling that is
quadratic in the lattice spacing. This describes the data $\beta \geq
6.0$ well (with $\chi^2/{\rm dof} = 0.178$). Even $\beta=4.5$ only
deviates from this line by 7\%, which backs up our previous statements
on scaling and the applicability of the perturbative lines of constant
physics (used to maintain constant $x,y$ as we varied $\beta$). In
addition, this fit suggests that residual lattice spacing corrections
are indeed very small at $\beta=18$, being around 2\% in this case.

\begin{figure}
\epsfig{file=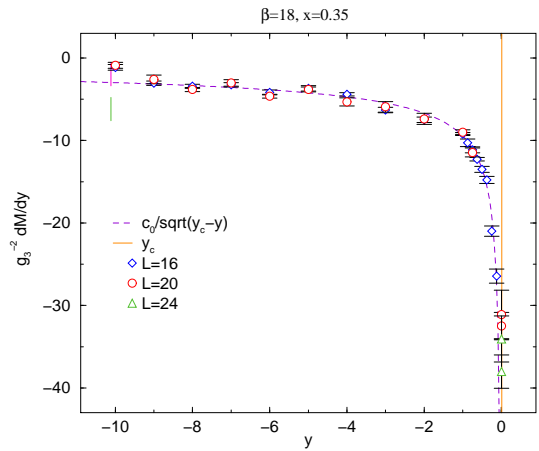,width=7cm}
\caption{\label{fig:deriv}
The derivative of the monopole free energy for intermediate
system sizes for fixed $x$. Also shown is the a semiclassically
inspired fit to the data.}
\end{figure}

In the region of intermediate volumes, when the twisted lattice
supports a single monopole, we may attempt to measure the mass
directly, to test the applicability of the
semiclassical results to fully quantised excitations. We have two
methods of approaching this. Less prone to statistical uncertainty is
to use measurements of the derivative of the mass $dM/dy$ over
a range in $y$ at fixed $x$. We make a `mean field' assumption that we
can describe this data using the formul{\ae} of
Section~\ref{sec:expectations}, allowing for a shift in the phase
transition by the substitution $y \to y - y_c$. Typical data, with such
a fit, are shown in Fig.~\ref{fig:deriv}. The mean field assumption fits
the data well, and from the coefficient $c_0$ we may extract a value
for the radiatively corrected 't Hooft function. We find $y_c$ to be
consistent with zero for $x=0.35$.

\begin{figure}
\epsfig{file=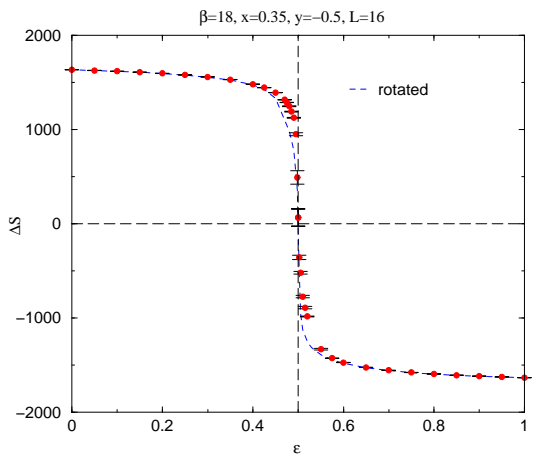,width=7cm}
\caption{\label{fig:prog_twist}
Measuring the monopole free energy by progressive twisting of
an intermediately--sized system. We show a 180 degree rotation of the
data to highlight the asymmetry.}
\end{figure}

Alternatively, we can measure the mass directly by the method of
progressive twisting for fixed $x,y$. We show such a calculation in
Fig.~\ref{fig:prog_twist}. The dominant error arises from the almost
complete cancellation of the areas under the curve either side of
$\varepsilon=0.5$. To illustrate this we plot also the same curve rotated
through $180^\circ$.

\begin{table}
\begin{tabular}{|l|l|l|l|l|l|l|}
\hline \hline
\multicolumn{1}{|c|}{$x$} & 
\multicolumn{1}{c|}{method} & 
\multicolumn{1}{c|}{$y$} & 
\multicolumn{1}{c|}{$M/g_3^2$} & 
\multicolumn{1}{c|}{$f(x)$} & 
\multicolumn{1}{c|}{$\beta$} &
\multicolumn{1}{c|}{$L$} \\
\hline
0.05 & deriv. & \multicolumn{2}{c|}{---} & 1.066 (11) & 18.0 & $16 - 20$ \\
0.35 & deriv. & \multicolumn{2}{c|}{---} & 1.257 (14) & 18.0 & $16 - 24$ \\
\hline
0.35 & prog. twist & $-10$ & 50.8 (14.4) & 1.07 (31) & 18.0 & $16$ \\
0.35 & prog. twist & $-3$ & 33.3 (5.5) &   1.28 (22) & 18.0 & $16$ \\
0.35 & prog. twist & $-1$ & 18.2 (4.7) &   1.21 (32) & 18.0 & $16$ \\
0.35 & prog. twist & $-0.5$ & 13.4 (1.9) & 1.26 (18) & 18.0 & $16$ \\
\hline
0.35 & prog. twist & $1$ & 1.4 (2.1) & --- & 18.0 & $16$ \\
0.35 & prog. twist & $10$ & 1.5 (1.2) & --- & 18.0 & $16$ \\
\hline
0.35 & prog. twist & $-0.124$ & 2.8 (1.6) & 0.53 (31) & 4.5 & $16$ \\
0.35 & prog. twist & $-0.124$ & 5.8 (1.8) & 1.10 (35) & 9.0 & $16$ \\
0.35 & prog. twist & $-0.124$ & 6.3 (2.4) & 1.19 (46) & 12.0 & $16$ \\
\hline \hline
\end{tabular}
\caption{\label{tab:mass}
Estimates of the monopole mass and the 't Hooft function.}
\end{table}

\begin{figure}
\epsfig{file=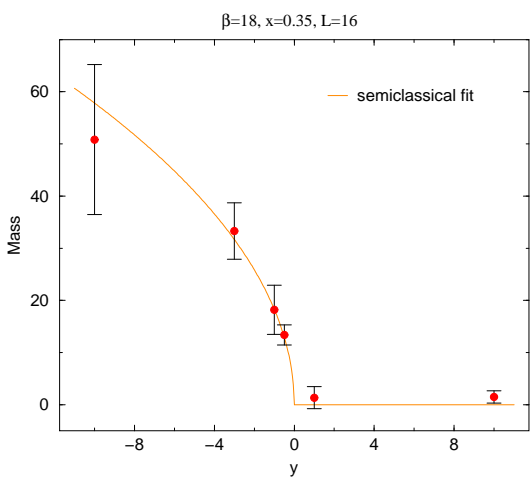,width=7cm}
\caption{
\label{fig:mass}
The monopole free energy as a function of $y$ for
intermediately sized system. A semiclassically inspired fit is show,
giving a value of the 't Hooft function.}
\end{figure}

We summarise these estimates of $f(x)$ in Table~\ref{tab:mass}, and in
Fig.~\ref{fig:mass} where we show a fit to different $y$ as per
Eq.~(\ref{eqn_sc_mass}). The masses and their derivatives behave much
as the semiclassical expectations. Similarly the 't Hooft function,
within the limits of our statistical errors, does not appear to differ
markedly due to radiative corrections. There is, however, a
considerable variation in the data at $y=-0.124$ as we change $\beta$,
and we may worry about systematic effects in our results. The first
source of these is discretisation effects. The majority of our
estimates are for $\beta=18.0$, and as we have argued above, the
residual lattice spacing effects are small here. The variation in
$\beta$ in the table is also in part due to a corresponding change in
the physical volume of the system, and we may ask whether all our
measurements are for `plateau' masses uncontaminated by the transient
small volume effects. We believe such biases to be small, especially
for the $y \leq -1$. As we vary $y$ in Fig.~\ref{fig:deriv} there is a
great change in the correlation lengths $\xi_{\{s,g\}}$ for fixed
volume. That the different {\it effective} volumes considered agree
suggests we are indeed seeing the intermediate plateau unaffected by
small $L$ transients. We are thus confident that our errors on these
estimates of $f(x)$ are accurate. The joint fit in Fig.~\ref{fig:mass}
yields $f(0.35) = 1.23 \, (12)$.

\subsection{Intermediate and large volumes}

\begin{figure}
\epsfig{file=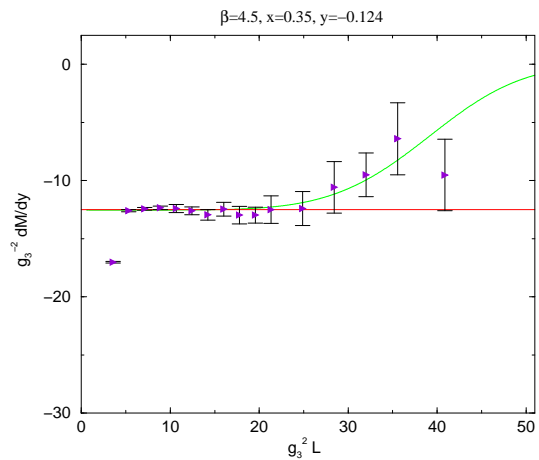,width=7cm}
\caption{\label{fig:screening}
The derivative of the monopole
free energy as a function of lattice size at $\beta = 4.5, x = 0.35, y
= -0.124$. Also shown are fits assuming no screening and a dilute gas
screening picture.}
\end{figure}

The large to intermediate system size data for the derivative
of the free energy are shown in Fig.~\ref{fig:screening}.
For intermediate system size it is clear that the
data is well represented by a constant independent of the lattice
size, and we use such a hypothesis:
\begin{equation}
\frac{1}{g_3^2} \frac{\partial}{\partial y} \frac{M}{g_3^2} = 
c_0 
\label{equ:const-fit}
\end{equation}
where we expect the parameter $c_0$ to be $p_0/g_3^2$. We show such
fits in Table~\ref{tab:const-fit}. Our method is to begin with a low
upper limit for the fitting range, and to then increase this,
including progressively more data in the fit. The $\chi^2$ per degree
of freedom and $Q$ (if our fitted form is the correct model, the
probability that our data could have arisen as random fluctuations
around that model) remain (very) acceptable up to $L \sim 40$. It is
clear that beyond this the fits become unacceptable: the behaviour has
changed as a consequence of screening.

\begin{table}
\begin{tabular}{|c|c|c|*{3}{r@{.}l|}}
\hline \hline
{\small $L_{\rm low}$} & 
{\small $L_{\rm high}$} & 
{\small $N_{\rm dof}$} &
\multicolumn{2}{c|}{$c_0$} &
\multicolumn{2}{c|}{$\chi^2/{\rm dof}$} & 
\multicolumn{2}{c|}{$Q$} \\
\hline
6 & 32 & 10 & 12&431 (58) & 0&686 & 0&722 \\
6 & 36 & 11 & 12&426 (58) & 1&106 & 0&353 \\
6 & 40 & 12 & 12&421 (58) & 1&704 & 0&066 \\
6 & 46 & 13 & 12&419 (58) & 1&813 & 0&040 \\
\hline
10 & 32 &  8 & 12&705 (82) & 0&772 & 0&628 \\
10 & 36 &  9 & 12&689 (82) & 1&229 & 0&272 \\
10 & 40 & 10 & 12&677 (82) & 1&873 & 0&044 \\
10 & 46 & 11 & 12&671 (82) & 1&976 & 0&027 \\
\hline \hline
\end{tabular}
\caption{\label{tab:const-fit}
Fitting the derivative of the
  free energy with an unscreened ansatz in Eq.~(\ref{equ:const-fit}).}
\end{table}

We can attempt to describe the screening by fitting over a similar
range using a fitting ansatz suggested by the dilute gas model:
\begin{equation}
\frac{1}{g_3^2} \frac{\partial}{\partial y} \frac{M}{g_3^2} = 
4 c_0 c_1 (ag_3^2L)^3
\frac{e^{-2 c_1 (a g_3^2 L)^3}}{1-e^{-4 c_1 (a g_3^2 L)^3}}
\label{equ:screen-fit}
\end{equation}
where $c_0$ is as before, and $c_1 = \nu_0/g_3^6$. We show
such fits over similar ranges in Table~\ref{tab:screen-fit}. For
intermediate $L$ the fits are similar to those obtained using just a
constant description. As data from larger systems is included,
however, we see that the ansatz now remains good. A comparison of the
two fits is plotted in Fig.~\ref{fig:screening}.

\begin{table}[t]
\begin{tabular}{|c|c|c|*{4}{r@{.}l|}}
\hline \hline
{\small $L_{\rm low}$} & 
{\small $L_{\rm high}$} & 
{\small $N_{\rm dof}$} &
\multicolumn{2}{c|}{$c_0$} &
\multicolumn{2}{c|}{$c_1 (\times 10^5$)} &
\multicolumn{2}{c|}{$\chi^2/{\rm dof}$} & 
\multicolumn{2}{c|}{$Q$} \\
\hline
6 & 32 &  9 & 12&499 (44) & 0&151 (93) & 0&938 & 0&490 \\
6 & 36 & 10 & 12&500 (44) & 0&187 (40) & 0&867 & 0&564 \\
6 & 40 & 11 & 12&502 (44) & 0&202 (35) & 0&826 & 0&614 \\
\hline
10 & 32 &  7 & 12&445 (83) & 1&36 (106) & 0&816 & 0&582 \\
10 & 36 &  8 & 12&451 (83) & 1&82 (47) & 0&753 & 0&675 \\
10 & 40 &  9 & 12&454 (83) & 1&99 (37) & 0&722 & 0&689 \\
10 & 46 & 10 & 12&449 (83) & 1&79 (35) & 0&773 & 0&655 \\
\hline \hline
\end{tabular}
\caption{\label{tab:screen-fit}
Fitting the derivative of the
  free energy with a dilute gas screening ansatz in 
Eq.~((\ref{equ:screen-fit}).}
\end{table}

We may calculate from $c_1$ the mean density of monopoles,
$\nu_0 / g_3^6 = 1.3 \, (3) \times 10^{-5}$, which makes their mean
separation
\begin{equation}
g_3^2 D = 42.6 \, (3.3) 
\end{equation}
or, in lattice units at $\beta = 4.5$, $\hat{D} = 47.8 \, (3.7)$. From
this it is clear that we have not got the lattice volume necessary to
see a complete screening of the free energy at $L \gg \hat{D}$.  We
cannot therefore completely rule out from our data the possibility
that $\Delta F$ remains finite even in the infinite volume limit.

As was seen for small system sizes, the plateau values at least are
heavily influenced by discretisation effects at $\beta = 4.5$. To
perform a scaling study of the screening mechanism is beyond our
current means. Nonetheless, for a demonstration of the mechanism such
effects are immaterial and do not affect the qualitative arguments.

Note also that no attempt has been made to estimate here the
systematic errors in the monopole density. To do so would require a
comparison of different screening hypotheses and fit functions,
something that the data is, unfortunately, not accurate enough to
address satisfactorily.

\section{Conclusions}
\label{sec:conclusions}

In this paper, we have used a fully non-perturbative technique to
measure the free energy of a 't~Hooft--Polyakov monopole in the
three--dimensional Georgi--Glashow model. This was achieved by
simulating systems with two different boundary conditions, both of
which are periodic up to symmetries of the Lagrangian. This preserves
the lattice translation invariance of the system and therefore makes
sure there are no boundary effects.

We found that in the Higgs phase, the free energy reached a constant
value at intermediate volumes, which shows that it is associated with
a localised object. This is the quantum analogue of the
't~Hooft--Polyakov monopole. We measure its mass by two different methods,
and find it compatible with semiclassical expectations. `Mean field'
application of the classical relations appears successful, and we can make
estimates of the quantum corrected 't Hooft function. Our best
estimates are $f(0.05) = 1.066 \, (11)$ and $f(0.35) = 1.257 \, (14)$
from the derivative of the mass with respect to $y$, and $f(0.35) =
1.23 \, (12)$ by the method of progressive twisting. These estimates
are both self--consistent, and in agreement with the classical
variation $f(x) \simeq 1 + x$ for small $x$
\cite{goddard78}, indicating that radiative
corrections are small.

When the volume increased above a certain critical
value, however, the free energy started to approach zero. This is
consistent with an analytical calculation within the dilute monopole
gas approximation, which predicts that the free energy vanishes in the
infinite volume limit at any values of the couplings.

In the dual superconductor picture, the vanishing monopole free energy
implies confinement, and therefore our results are numerical evidence
for Polyakov's prediction that the Higgs phase of this theory is
confining. Furthermore, if the monopole free energy vanishes
everywhere, it cannot be used as an order parameter, and therefore our
results strongly support the conjecture that the confining and Higgs
phases are analytically connected to one another.

Neither, of course, can the monopole mass measured from the plateau in
the free energy for intermediate system sizes act to distinguish the
phases. It is non-zero in the deep Higgs phase and zero in the deep
symmetric phase. This plateau does not exist, however, everywhere in
the phase diagram, notably near the transition line itself. The mean
monopole separation there will be comparable to the core size and no
plateau would be observed. Thus the `mass' is ill--defined and cannot
serve as an order parameter.

Our findings suggest a straightforward generalisation to other cases.
In a Euclidean formulation in any number of dimensions, any point-like
topological defect that has finite action, will always have a non-zero
density at any non-zero temperature. This means that these objects
always have a zero free energy. An extended topological defect, such
as a string or a domain wall, is, however, either a closed loop,
surface etc., in which case it does not contribute to the global
properties of the systems, or it has an infinite action. In the latter
case, the fluctuations cannot generate them, and their free energy
remains non-zero even in the infinite volume limit. Because the free
energy can be used as an order parameter, this suggests that models
with extended topological defects always have a true phase transition
rather than a smooth crossover.  This question will be studied further
in a future publication.

\begin{acknowledgments}
We would like to thank Mikko Laine and Kari
Rummukainen
for useful
discussions.  This work was supported by PPARC(UK) and by the ESF
COSLAB Programme. The computational work was carried out on the
U.K. Computational Cosmology Consortium COSMOS Origin2000
supercomputer.
\end{acknowledgments}

%
%

%
%
%
%
%


\begin{thebibliography}{99}

\bibitem{'tHooft:1974qc}
G.~'t Hooft,
Nucl.\ Phys.\  {B 79} (1974) 276.

\bibitem{Polyakov:1974ek}
A.~M.~Polyakov,
JETP Lett.\  {20} (1974) 194.

\bibitem{Fradkin:1979dv}
E.~Fradkin and S.~H.~Shenker,
Phys.\ Rev.\  {D 19} (1979) 3682.

\bibitem{Nadkarni:1988pb}
S.~Nadkarni,
Phys.\ Rev.\ Lett.\  {60} (1988) 491;
Nucl. Phys. B 334 (1990) 559.

\bibitem{Hart:1997ac}
A.~Hart, O.~Philipsen, J.~D.~Stack and M.~Teper,
Phys.\ Lett.\ B {396} (1997) 217
\eprint{[hep-lat/9612021]}.

\bibitem{Kajantie:1997tt}
K.~Kajantie, M.~Laine, K.~Rummukainen and M.~Shaposhnikov,
Nucl.\ Phys.\ B {503} (1997) 357
\eprint{[hep-ph/9704416]}.

\bibitem{Kleinert:1982dz}
H.~Kleinert,
Lett.\ Nuovo Cim.\  {35} (1982) 405.

\bibitem{Kovner:1991pz}
A.~Kovner, B.~Rosenstein and D.~Eliezer,
Nucl.\ Phys.\  {B 350} (1991) 325.

\bibitem{Kajantie:1999zn}
K.~Kajantie, M.~Laine, T.~Neuhaus, J.~Peisa, A.~Rajantie and K.~Rummukainen,
Nucl.\ Phys.\  {B 546} (1999) 351
\eprint{[hep-ph/9809334]}.

\bibitem{Davis:2000kv}
A.~C.~Davis, T.~W.~B.~Kibble, A.~Rajantie and H.~Shanahan,
JHEP {0011} (2000) 010
\eprint{[hep-lat/0009037]}.

\bibitem{Polyakov:1977fu}
A.~M.~Polyakov,
Nucl.\ Phys.\ B {120} (1977) 429.

\bibitem{polyakov87}
A.~M.~Polyakov,
Gauge Fields and Strings (Harwood, 1987).

\bibitem{Mandelstam:1976pi}
G.~'t~Hooft, in ``High Energy Physics'', EPS International Conference,
Palermo 1975, ed. A.~Zichichi;
S.~Mandelstam,
Phys.\ Rept.\  {23} (1976) 245.

\bibitem{Kajantie:1996dw}
K.~Kajantie, M.~Laine, K.~Rummukainen and M.~Shaposhnikov,
Nucl.\ Phys.\ B {458} (1996) 90
\eprint{[hep-ph/9508379]}.

\bibitem{hart99}
A. Hart and O. Philipsen,
Nucl. Phys. B 572 (2000) 243
\eprint{[hep-lat/9908041]};
A. Hart, M. Laine and O. Philipsen. 
Nucl. Phys. B 586 (2000) 443
\eprint{[hep-ph/0004060]};
Phys. Lett. B 505 (2001) 141 
\eprint{[hep-lat/0010008]}.

\bibitem{'tHooft:1981ht}
G.~'t Hooft,
Nucl.\ Phys.\  {B 190} (1981) 455.

\bibitem{Smit:1994vt}
J.~Smit and A.~J.~van der Sijs,
Nucl. Phys. B 422 (1994) 349
\eprint{[hep-lat/9312087]}.

\bibitem{DelDebbio:1995sx}
L.~Del Debbio, A.~Di Giacomo and G.~Paffuti,
Phys.\ Lett.\  {B 349} (1995) 513
\eprint{[hep-lat/9403013]};
%
A.~Di Giacomo, B.~Lucini, L.~Montesi and G.~Paffuti,
Phys.\ Rev.\  {D 61} (2000) 034503
\eprint{[hep-lat/9906024]}.

\bibitem{Frohlich:1999wq}
J.~Fr\"ohlich and P.~A.~Marchetti,
Nucl. Phys. B 551 (1999) 770
\eprint{[hep-lat/9812004]}.

\bibitem{Cea:2000zr}
P.~Cea and L.~Cosmai,
Phys.\ Rev.\ D {62} (2000) 094510
\eprint{[hep-lat/0006007]}.

\bibitem{farakos94}
K.~Farakos, K.~Kajantie, K.~Rummukainen and M.~Shaposhnikov,
Nucl.\ Phys.\  {B 442} (1995) 317
\eprint{[hep-lat/9412091]}.

\bibitem{Laine:1995ag}
M.~Laine,
Nucl.\ Phys.\ B {451} (1995) 484
\eprint{[hep-lat/9504001]}.

\bibitem{laine97}
M.~Laine and A.~Rajantie,
Nucl.\ Phys.\  {B 513} (1998) 471
\eprint{[hep-lat/9705003]}.

\bibitem{Kronfeld:1991qu}
A.~S.~Kronfeld and U.~J.~Wiese,
Nucl. Phys. B 357 (1991) 521;
Nucl. Phys. B 401 (1993) 190
\eprint{[hep-lat/9210008]}.

\bibitem{'tHooft:1978hy}
G.~'t Hooft,
Nucl.\ Phys.\  {B 138} (1978) 1.

\bibitem{Kovacs:2000sy}
T.~G.~Kovacs and E.~T.~Tomboulis,
Phys.\ Rev.\ Lett.\  {85} (2000) 704
\eprint{[hep-lat/0002004]}.

\bibitem{Hoelbling:2001su}
C.~Hoelbling, C.~Rebbi and V.~A.~Rubakov,
Phys.\ Rev.\ D {63} (2001) 034506
\eprint{[hep-lat/0003010]}.

\bibitem{hart00d}
A. Hart, B. Lucini, Z. Schram and M. Teper, 
JHEP 11 (2000) 043
\eprint{[hep-lat/0010010]}.

\bibitem{deForcrand:2001fi}
P.~de Forcrand, M.~D'Elia and M.~Pepe,
Phys.\ Rev.\ Lett.\  {86} (2001) 1438
\eprint{[hep-lat/0007034]}.

\bibitem{Chernodub:2001nz}
M.~N.~Chernodub, F.~V.~Gubarev, M.~I.~Polikarpov and V.~I.~Zakharov,
Phys.\ Lett.\ B {514} (2001) 88
\eprint{[hep-ph/0101012]}.

\bibitem{goddard78}
P.~Goddard and D.~I.~Olive,
Rept.\ Prog.\ Phys.\  {41} (1978) 1357.

\bibitem{hart00c}
A. Hart, B. Lucini, Z. Schram and M. Teper, 
JHEP 06 (2000) 040
\eprint{[hep-lat/0005010]}.

\bibitem{teper98}
M. Teper,
Phys. Rev. D 59 (1999) 014512
\eprint{[hep-lat/9804008]}.


\end{thebibliography}
\end{document}